\documentstyle[aps,prl,amsfonts]{revtex}
\input{epsf}
\begin{document}
\title {
  Lowest Landau level broadened by a Gau{ss}ian random potential with
  an arbitrary correlation length: An efficient continued-fraction
  approach
}
\author{
Markus B\"ohm,${}^1$ Kurt Broderix,${}^2$ and Hajo Leschke${}^1$}
\address{
${}^1$%
Institut f\"ur Theoretische Physik,
Universit\"at Erlangen-N\"urnberg,
Staudtstr.\ 7,
D-91058 Erlangen,
Germany}
\address{
${}^2$%
Institut f\"ur Theoretische Physik,
Universit\"at G\"ottingen,
Bunsenstr.\ 9,
D-37073 G\"ottingen,
Germany}
\date{\today}
\maketitle
\begin{abstract}
For an electron in the
plane subjected to a perpendicular constant magnetic field and a
homogeneous Gau{ss}ian random potential
with a Gau{ss}ian covariance function we approximate the averaged
density of states restricted to the lowest Landau level.  To this end,
we extrapolate the first 9 coefficients of the underlying continued
fraction consistently with the coefficients' high-order asymptotics.
We thus achieve the first reliable extension of Wegner's exact result
[Z.\ Phys.\ B {\bf 51}, 279 (1983)] for the delta-correlated case to
the physically more relevant case of a non-zero correlation length.
\end{abstract}

\narrowtext

Nearly ideal two-dimensional electronic structures have attracted
great attention for more than a decade not only because of their
varied and important applications, but also because of the discovery
of the quantum Hall effect\cite{review}.  For a microscopic
understanding of the occurring phenomena it is essential to know the
spectral properties of electrons confined to two dimensions under the
influence of a perpendicular constant magnetic field taking into
account the presence of disorder.

A commonly studied minimal model is that of non-interacting electrons
which is characterized by the one-electron Hamiltonian given by the
Schr\"odinger operator
\begin{equation}
\label{H}
\hat{H} := \hat{K} + \hat{V}
\end{equation}
\begin{equation}
\hat{K} :=
\frac{1}{2m}\left(
\frac{\hbar}{\text{i}} \frac{\partial}{\partial x_1} -
\frac{\text{e}B}{2}x_2
\right)^2
+
\frac{1}{2m}\left(
\frac{\hbar}{\text{i}}
\frac{\partial}{\partial x_2} +
\frac{\text{e}B}{2}x_1
\right)^2
{}.
\end{equation}
Here $x:=(x_1,x_2)$ are Cartesian coordinates of the Euclidean plane
${\Bbb R}^2$, $\hbar$ is Planck's constant, e is the elementary
charge, $m$ is the (effective) mass of the (spinless) electron, and
$B>0$ the strength of a perpendicular magnetic field.  The static
random potential $V$ is added to mimic the interaction with quenched
disorder.  Its probability distribution is assumed to be Gau{ss}ian
with zero mean and Gau{ss}ian covariance, that is
\begin{equation}
\label{covariance}
\overline{V(x)}=0
,
\quad
\overline{V(x)V(x')} = \sigma^2
\exp\!\left[-\frac{(x-x')^2}{2\lambda^2}\right]
{}.
\end{equation}
The overbar denotes averaging with respect to the probability
distribution.  Finally, $\sigma>0$ is the strength and $\lambda>0$ the
correlation length of the fluctuations of the potential.

The spectral resolution of the unperturbed Hamiltonian $\hat{K}$ reads
$\hat{K}=\sum_{n=0}^\infty \varepsilon_n \hat{E}_n$, where the
eigenvalue $\varepsilon_n := (2n+1) \frac{\hbar \text{e} B}{2 m}$ is
called the $n$-th Landau level and $\hat{E}_n$ denotes the
corresponding eigenprojector.  With increasing $B$ the magnetic length
$l:=\sqrt{\hbar/(\text{e}B)}$ decreases, while the degeneracy $\langle
x | \hat{E}_n | x \rangle = (2\pi l^2)^{-1}$ (per area) of each Landau
level and the distance between successive levels increase.  Hence, for
a fixed concentration of electrons, sufficiently high fields, and low
temperatures it is reasonable to simplify the model by restricting its
Hamiltonian (\ref{H}) to the (still infinite dimensional) eigenspace
$\hat{E}_0L^2({\Bbb R}^2)$ of the lowest Landau level.  This has been
done, for example, in the important work of Wegner \cite{wegner},
where the averaged restricted density of states (per area)
\begin{equation}
\label{dos}
\varrho(\varepsilon) :=
\overline{ \langle y|\hat{E}_0 \delta(\varepsilon
- \hat{E}_0 \hat{H} \hat{E}_0) \hat{E}_0 |y \rangle }
\end{equation}
has been calculated exactly in the delta-correlated limit $\lambda
\downarrow 0$, $\sigma \to \infty$, $\lambda^2 \sigma^2 =
\mbox{const}$.

The purpose of the present Letter is twofold.  One goal is to present
what we think is an accurate extension of Wegner's result to arbitrary
values of $\sigma$ and $\lambda$.  Clearly, this is of physical
interest because in the high-field limit the actual correlation length
is no longer small in comparison with the magnetic length.  The other
goal is to demonstrate thereby that the power of the well-known
continued-fraction approach to spectral densities of non-trivial
quantum problems \cite{cookbook,czycholl} can considerably be enhanced
in cases, where one has an {\em a-priori} information about their
asymptotic high-frequency decay.

To get rid of physical dimensions we write $\varrho$ in the
standardized form
\begin{equation}
\label{dimlessdos}
\varrho(\varepsilon) =:
\frac{\sqrt{1+(l/\lambda)^2}}{2\pi l^2 \sigma}
\;
W\!\left(
\frac{\lambda^2}{l^2},
\frac{(\varepsilon-\varepsilon_0)}{\sigma}\sqrt{1+(l/\lambda)^2}
\right)
{}.
\end{equation}
In this way we have introduced a one-parameter family of even
probability densities on the real line ${\Bbb R}$ with normalized
second moment
\begin{equation}
W(a,u) = W(a,-u) \ge 0
\end{equation}
\begin{equation}\label{normalization}
\int_{\Bbb R} \! \text{d} u \, W(a,u) = 1
= \int_{\Bbb R} \! \text{d} u \, W(a,u) \, u^2
{}.
\end{equation}
Here we have used Eq.~(7) of \cite{jphysa} to normalize the second
moment.

In this notation Wegner's result for the delta-correlated limit reads
\cite{wegner,brezin}
\begin{equation}
\label{wegnerdos}
W(0,u) =
\frac{ 2 \pi^{-3/2} \exp(u^2) }{1+ \left[
2 \pi^{-1/2}
\int_0^u \text{d}\xi \, \exp(\xi^2)
\right]^2}
{}.
\end{equation}
For the other extreme of the spatial extent of correlations, namely
the constantly correlated case $\lambda = \infty$, one simply has
\cite{zphys} the Gau{ss}ian
\begin{equation}
\label{constantdos}
W(\infty,u) =
(2\pi)^{-1/2}\exp\!\left(-u^2/2\right)
{}.
\end{equation}
For intermediate values of $a$ no exact expression for $W(a,u)$ is
known.  What is exactly known, however, is the fact that $W(a,u)$
falls off for sufficiently large $|u|$ like a Gau{ss}ian.  More
precisely, by Eqs. (17) and (15) of \cite{jphysa} (see also
\cite{apel})
\begin{equation}
\label{decay}
\lim_{u \to \pm \infty} \frac{1}{u^2} \ln( W(a,u)) =
-\frac{a+2}{2a+2}
{}.
\end{equation}

In the sequel we will design approximations to the Stieltjes transform
\begin{equation}
R(a,z) :=
\int_{\Bbb R} \! \text{d}u \, \frac{W(a,u)}{z - \text{i}u}
, \quad
\mbox{Re} z > 0
,
\end{equation}
of the standardized density of states $W$ which in turn will yield
approximations to $W$ by means of the inversion formula
\begin{equation}
\label{inversion}
W(a,u) = \frac{1}{\pi} \lim_{v \downarrow 0}
\mbox{Re}\left[
R(a,v+\text{i}u)
\right]
{}.
\end{equation}
According to Stieltjes' classical theory, see for example
\cite{perron,wall}, $R$ can be expanded into a Jacobi-type continued
fraction
\begin{equation}
R(a,z) = \mathop{\text{K}}_{j=1}^\infty (\frac{r_j(a)}{z})
, \quad
r_j(a) \ge 0
{}.
\end{equation}
Here we are using the notation
\begin{equation}
\mathop{\text{K}}_{j=1}^\infty (\frac{\Delta_j}{z})
:=
\frac{\displaystyle 1}{\displaystyle z +
\frac{\displaystyle \Delta_1}{\displaystyle z +
\frac{\displaystyle \Delta_2}{\displaystyle z +
\ddots
}
}
}
\end{equation}
for the continued fraction with coefficients $\Delta_1, \Delta_2,
\ldots$ and variable $z$.

To derive the continued-fraction coefficients $\{ r_j(0) \}$ and $\{
r_j(\infty) \}$ corresponding to (\ref{wegnerdos}) and
(\ref{constantdos}), respectively, we use the identity
\begin{equation}
\label{terminator}
\mathop{\text{K}}_{j=1}^\infty (\frac{\beta + \gamma j}{z})
=
\frac{ D_{-(\beta/\gamma)-1} ( \gamma^{-1/2} z ) }{
\gamma^{1/2} D_{-\beta/\gamma} ( \gamma^{-1/2} z )}
=:
T(\beta,\gamma,z)
,
\end{equation}
valid if $\gamma>0$, $\beta+\gamma>0$, and $\mbox{Re}z>0$.  Here
$D_\nu$ denotes Whittaker's parabolic cylinder function with index
$\nu$, see Section 8.1 of \cite{magnus}.  The identity follows by
iteration from the observation that the rhs of (\ref{terminator})
obeys
\begin{equation}
\label{recurrence}
T(\beta,\gamma,z) = [z+(\beta+\gamma)T(\beta+\gamma,\gamma,z)]^{-1}
\end{equation}
which itself is a consequence of the first of the four recurrence
relations for the $D_\nu$'s in Section 8.1.3 of \cite{magnus}.  In
fact, Eq.~(\ref{terminator}) is equivalent to Eq.~(14) in \S 50 of
\cite{perron}.  With the help of (\ref{inversion}) one now checks that
\begin{equation}
r_j(0) = \frac{1}{2} + \frac{1}{2} j
, \quad
r_j(\infty) = j
,
\end{equation}
for all $j \ge 1$.

According to \cite{growth-rate} the asymptotic behavior (\ref{decay})
implies the following asymptotic linear growth for the
continued-fraction coefficients
\begin{equation}
\label{asympty}
\lim_{j\to\infty}
\frac{r_j(a)}{j}
 =
\frac{a+1}{a+2}
{}.
\end{equation}
It is natural to match this linear growth to the first $J < \infty$
coefficients $r_1(a)$, \ldots, $r_J(a)$ to construct the announced
approximations $R^{(J)}(a,z)$ to $R(a,z)$ by means of
\begin{equation}
R^{(J)}(a,z) :=
\mathop{\text{K}}_{j=1}^\infty (\frac{r^{(J)}_j(a)}{z}),
\end{equation}
where
\begin{equation}
\label{match}
r_j^{(J)}(a) :=
\left\{
\begin{array}{ll}
r_j(a) & \quad \mbox{for} \quad j \le J \\
r_J(a) + \frac{a+1}{a+2} (j-J) & \quad \mbox{for} \quad j > J
\end{array}
\right.
{}.
\end{equation}
Not surprisingly, the approximation $R^{(J)}$ to $R$ results in an
approximation $W^{(J)}$ to $W$ with the same Gaussian fall-off
(\ref{decay}) in the tails.  This can be seen as follows.  By virtue
of (\ref{terminator}) the approximation $R^{(J)}$ can be expressed as
a terminating continued fraction
\begin{equation}
R^{(J)}(a,z) =
\frac{\displaystyle 1}{\displaystyle z +
\frac{\displaystyle r_1(a)}{\displaystyle z
+_{\ddots\displaystyle +
\frac{\displaystyle r_{J-1}(a)}{\displaystyle z +
r_J(a)T(r_J(a),{\textstyle\frac{a+1}{a+2}},z)
}
}
}
}.
\end{equation}
In view of (\ref{inversion}) it is therefore sufficient to show
\begin{equation}
\label{t-prop-re}
\lim_{u \to \pm\infty} \frac{1}{u^2} \ln(\mbox{Re}[T(\beta,\gamma,iu)]) =
-\frac{1}{2\gamma}
\end{equation}
and
\begin{equation}
\label{t-prop-im}
\lim_{u \to\pm\infty} \mbox{Im}[T(\beta,\gamma,iu)] = 0,
\end{equation}
because these asymptotic properties are conserved under the
(repeated) substitution
\begin{equation}
T(\beta,\gamma,iu) \mapsto [iu + \alpha T(\beta,\gamma,iu)]^{-1}
\end{equation}
for any $\alpha>0$.  The validity of (\ref{t-prop-re}) und
(\ref{t-prop-im}) can be deduced from an asymptotic evaluation
\cite{bender} of the following Riccati differential equation
\begin{equation}
\gamma \frac{d}{du} T(\beta,\gamma,iu) =
i\beta \left[T(\beta,\gamma,iu)\right]^2 - u T(\beta,\gamma,iu) - i
\end{equation}
which itself follows from the aforementioned recurrence relations. To
summarize, Eq.~(\ref{terminator}) adds a two-parameter family of
terminators for affine-linear extrapolations to the toolkit of
\cite{cookbook}.

In order to compute the first $J$ continued-fraction coefficients
$r_1(a)$, \ldots, $r_J(a)$, we employ the well-known one-to-one
correspondence \cite{perron,wall,cookbook} to the first $J$ even
moments $M_2(a)$, \ldots, $M_{2J}(a)$ of the standardized density of
states.

According to (\ref{dos}) and (\ref{dimlessdos}) one has for the latter
\begin{eqnarray}
\label{momdef}
M_{2j}(\lambda^2/l^2) :=
\int_{\Bbb R} \! \text{d}u\,
W(\lambda^2/l^2,u)u^{2j}
\nonumber
\\ =
2\pi l^2 \left(\frac{1+l^2/\lambda^2}{\sigma^2}\right)^j
\overline{ \langle y|(\hat{E}_0 \hat{V} \hat{E}_0)^{2j}|y \rangle }
\end{eqnarray}
if $j \ge 1$.  We now substitute $\hat{V}=\int_{{\Bbb R}^2}\text{d}^2x
\, V(x) | x \rangle \langle x |$ into the rhs of (\ref{momdef}) and
use the standard reduction formula
\begin{equation}
\overline{\prod_{k=1}^{2j}V(x(k))} =
\sum_{k=1}^{(2j-1)!!} \prod_{s=1}^j
\overline{V(x({P_k(2s-1)}))V(x({P_k(2s)}))}
\end{equation}
for the average of a product of $2j$ jointly Gau{ss}ian random
variables with zero mean.  Here $x(1),\ldots,x(2j) \in {\Bbb R}^2$ are
$2j$ points of the plane and $P_k$ denotes the $k$-th of the
$(2j-1)!!$ permutations of the first $2j$ natural numbers which lead
to different sets $\left\{ \{ P_k(1),P_k(2) \} , \ldots ,\{
P_k(2j-1),P_k(2j) \} \right\}$ of $j$ pairs $\{ P_k(2s-1),P_k(2s) \}$.
Since the covariance function (\ref{covariance}) of the random
potential $V$ and the position representation
\begin{equation}
 \langle x|\hat{E}_0|x' \rangle  =
\frac{1}{2\pi l^2}
\exp\!\left[
\frac{\text{i}}{2 l^2}(x_1x_2'-x_2x_1') - \frac{1}{4l^2}(x-x')^2
\right]
\end{equation}
of the eigenprojector corresponding to the lowest Landau level are both
Gau{ss}ian, one ends up with a sum over $4j$-dimensional Gau{ss}ian
integrals which can be performed to yield
\begin{equation}
\label{momsum}
M_{2j}(a) =
\Big(1+\frac{1}{a}\Big)^j
\:
\sum_{k=1}^{(2j-1)!!} \frac{1}{\det\left(A_k(a)\right)}
,
\quad a>0.
\end{equation}
Here the $2j\times 2j$-matrix $A_k(a)$ is defined through its
entries in terms of the Kronecker delta
\begin{equation}
\left(A_k(a)\right)_{\mu,\nu} :=
\Big(1+\frac{1}{a}\Big) \delta_{\mu,\nu} - \delta_{\mu+1,\nu} -
\frac{1}{a}\sum_{s=1}^j \left(
\delta_{\mu,P_k(2s-1)} \delta_{\nu,P_k(2s)}+
 \delta_{\nu,P_k(2s-1)} \delta_{\mu,P_k(2s)}
\right)
{}.
\end{equation}

We have computed the sum (\ref{momsum}) for $j=1, 2, \ldots, 6$ using
the {\sc Axiom} symbolic system \cite{axiom} for general $a$.  The
results for $M_{2j}(a)$, $j=1,2,3,4$, are given by the following
rational functions
\begin{equation}
\begin{array}{c}
M_{2}(b-1) = 1
\\[1ex]
M_{4}(b-1) = (3 b^2+2) / (b^2+1)
\\[1ex]
M_{6}(b-1) =
({{{{15}  {b^6}}+{{75}  {b^4}}+{{102}
{b^2}}+{30}}) / ({{b^6}+{6  {b^4}}+{{11}  {b^2}}+6} })
\\[1ex]
M_{8}(b-1) =
(
{{105} {b^{24}}}
+{{2835} {b^{22}}}
+{{32830} {b^{20}}}
+{{213697}  {b^{18}}}
+{{861130}  {b^{16}}}
+{{2231807}  {b^{14}}}\\
+{{3750338}  {b^{12}}}
+{{4038543}  {b^{10}}}
+{{2717117}  {b^8}}
+{{1105510}  {b^6}}
+{{261216}  {b^4}}
+{{32792}  {b^2}}
+{1680}
)/
\\
(
{b^{24}}
+{{29}  {b^{22}}}
+{{365}  {b^{20}}}
+{{2620}  {b^{18}}}
+{{11854}  {b^{16}}}
+{{35276}  {b^{14}}}
+{{69974}  {b^{12}}}
+{{91906}  {b^{10}}}  \\
+{{78025}  {b^8}}
+{{41015}  {b^6}}
+{{12461} {b^4}}
+{{1954}  {b^2}}
+{120}
).
\end{array}
\end{equation}
Since the expressions for $M_{10}(a)$ and $M_{12}(a)$ are extremely
lengthy, we only give, as an example, their values for $a=1$:
\begin{equation}
\begin{array}{c}
M_{10}(1)=
\frac{158\,659\,605\,940\,126\,452\,841}%
            {294\,310\,802\,651\,335\,470} \\[1ex]
M_{12}(1)=
\frac{388\,336\,271\,072\,847\,928\,549\,926\,597\,%
113\,071\,401\,088\,677\,997\,478\,405\,727\,555\,223\,031}%
{82\,363\,680\,790\,265\,452\,914\,044\,225\,729\,941\,%
466\,953\,642\,191\,484\,142\,275\,602\,750}.
\end{array}
\end{equation}
The moments $M_{14}(a)$ and $M_{16}(a)$ have been computed exactly as
a reduced fraction for $a=\frac{1}{4},\frac{1}{2},1,2,4$ only, by
employing a C-program.  For the same set of values for $a$ we also
have computed $M_{18}(a)$, but only have been able to gather up the
$17!! \approx 3.4 \cdot 10^7$ terms occurring in (\ref{momsum}) to a
sum of approximately $10^4$ reduced fractions, the value of which has
been computed accurately by floating-point arithmetics.  For lack of
space we only show the first 35 digits for the example $a=1$:
\begin{eqnarray*}
M_{14}(1)=47657.946072630475536554559639613945\ldots \\
M_{16}(1)=545841.43592501744224295351679205436\ldots \\
M_{18}(1)=6980770.3795705620571551127542699202\ldots
\end{eqnarray*}

\begin{figure}[h]

\vskip-33mm
\epsfxsize98mm\centerline{\epsfbox{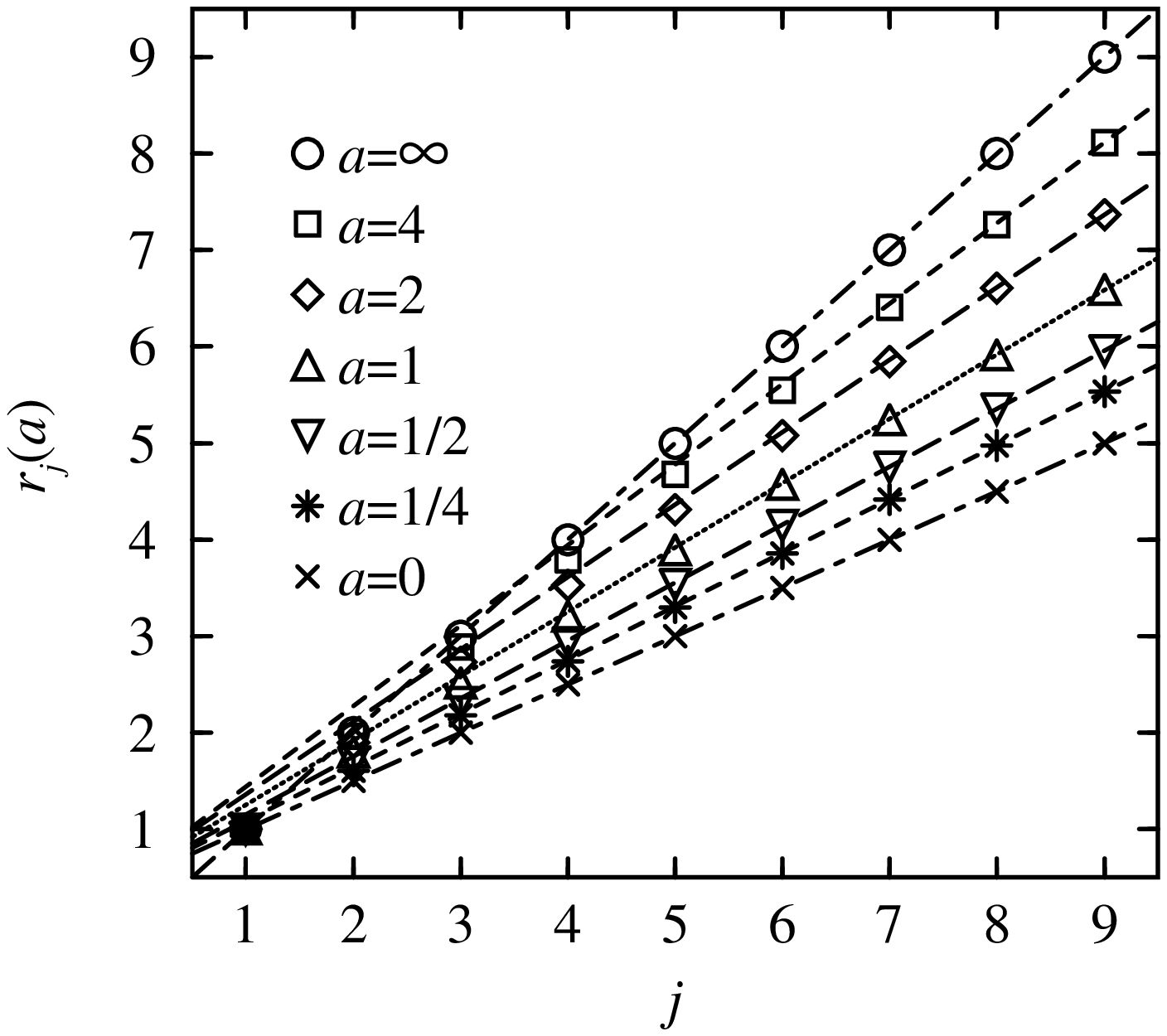}}

\vskip-28mm
\caption{\label{fig:delta}
  Symbols represent the first 9 continued-fraction coefficients
  $r_1(a),\dots,r_9(a)$ for different values of $a$.  Note that
  $r_1(a)=1$ for all $a$ by (\protect\ref{normalization}) The straight
  lines correspond to the asymptotics of $\{r_j^{(9)}(a)\}$ as defined
  in (\protect\ref{match}).
}
\end{figure}

{}From Fig.~\ref{fig:delta} it can be seen that the resulting
continued-fraction coefficients $r_1(a),\ldots,r_9(a)$ approach quite
rapidly their asymptotic behavior given in (\ref{asympty}).  For
example, the relative extrapolation error $|r_9^{(8)}(a) -
r_9(a)|/r_9(a)$ varies between $0.2\%$ for $a=4$ and $0.02\%$ for
$a=\frac{1}{4}$.

\begin{figure}[h]

\vskip-12mm
\hskip-25pt\epsfxsize94mm\centerline{\epsfbox{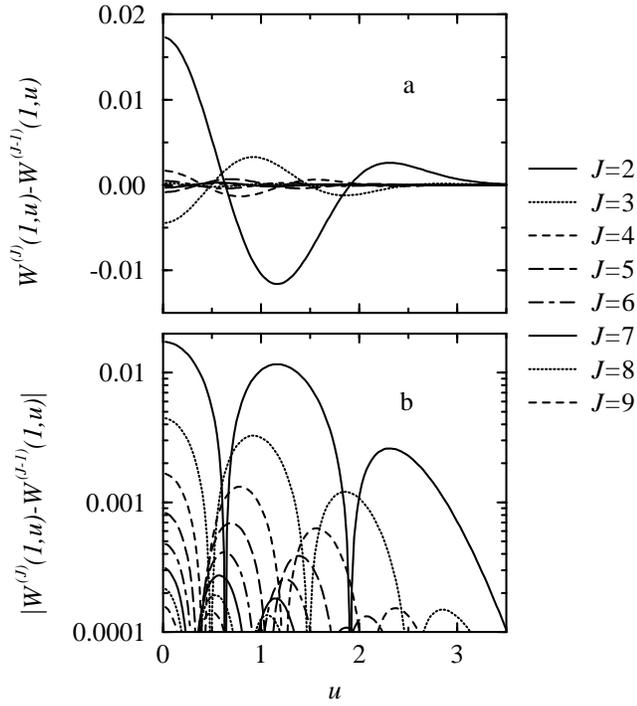}}

\vskip-15mm
\caption{\label{fig:convergence}
  Difference of two successive approximations to $W$ for $a=1$ as a
  function of $u$ given on a linear (a) and a logarithmic (b) scale.
}
\end{figure}

\begin{figure}[h]

\vskip-33mm
\epsfxsize98mm\centerline{\epsfbox{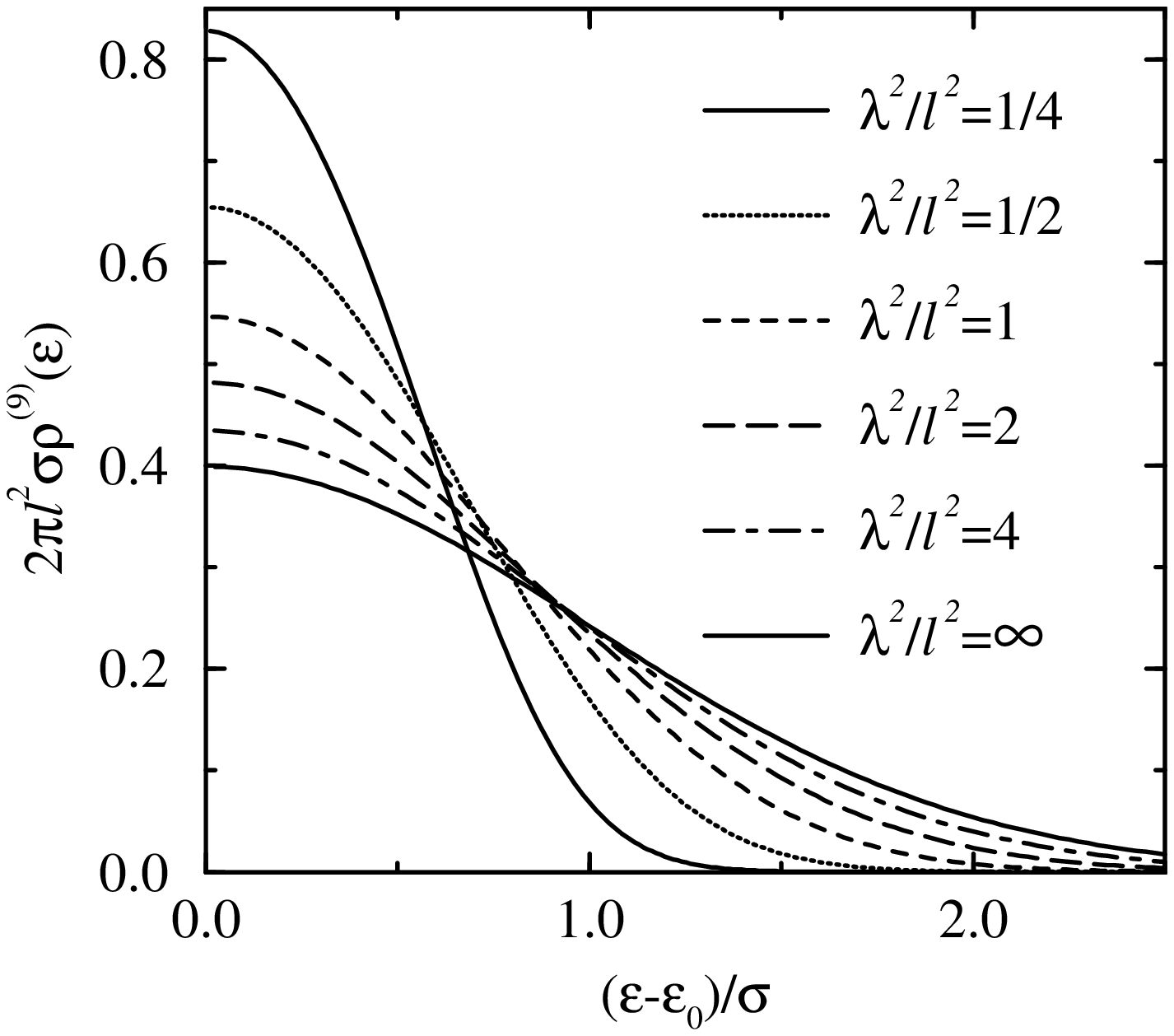}}

\vskip-28mm
\caption{\label{fig:dos}
  The approximation $\varrho^{(9)}$ to the averaged density of states
  $\varrho$ for different values of the correlation length $\lambda$
  as a function of the energy $\varepsilon$.
}
\end{figure}

In Fig.~\ref{fig:convergence} we demonstrate the convergence of the
resulting approximations $W^{(J)}(1,u)$ to $W(1,u)$ for increasing
$J=1,2,\ldots,9$.  The differences of two successive approximations,
$W^{(J)}(a,u)-W^{(J-1)}(a,u)$, seem to form a nearly alternating
sequence with geometric decrease for fixed $u$.  Both observations,
taken together, suggest rapid pointwise convergence of the sequence
$\{W^{(J)}\}$.  Therefore, we may safely conclude that $W^{(9)}$
constitutes a reliable approximation to $W$.  Figure \ref{fig:dos}
shows a plot of the corresponding approximation $\varrho^{(9)}$ to the
averaged density of states $\varrho$ for different $\lambda$.  Note
that, by construction, $\varrho^{(9)}$ is exact in the limiting cases
of a delta-correlated and a constantly correlated random potential;
more importantly, $\varepsilon\mapsto 2\pi
l^2\varrho^{(9)}(\varepsilon+\varepsilon_0)$ is an even probability
density with the same first $18$ moments and the same Gau{ss}ian
fall-off in the tails as the true density for general values of the
correlation length $\lambda$.


\begin{references}
\bibitem{review}
For review-type literature see
T. Ando, A. B. Fowler, and F. Stern,
  Rev. Mod. Phys. {\bf 54}, 437 (1982);
I. V. Kukushkin, S. V. Meshkov, and V. B. Timofeev,
  Usp. Fiz. Nauk. {\bf 155}, 219 (1988)
  [Sov. Phys. Usp. {\bf 31}, 511 (1988)];
G. Landwehr (ed),
  {\em High Magnetic Fields in Semiconductor Physics I, II, III}
  (Springer, Berlin, 1987, 1989, 1992).
M. Jan{\ss}en, O. Viehweger, U. Fastenrath, and J. Hajdu,
  {\em Introduction to the Theory of the Integer Quantum Hall Effect}
  (VCH, Weinheim, 1994)
\bibitem{cookbook}
V. S. Viswanath and G. M\"uller,
  {\em The Recursion Method --- Application to Many-Body Dynamics},
  Lecture Notes in Physics, Vol m23
  (Springer, Berlin, 1994), and references therein.
\bibitem{czycholl}
  Continued fractions were applied successfully to a lattice version of
  (\ref{H}) by
G. Czycholl and W. Ponischowski, Z. Phys. B {\bf 73}, 343 (1988).
\bibitem{jphysa}
K. Broderix, N. Heldt, and H. Leschke, J. Phys. A {\bf 24}, L825 (1991).
\bibitem{wegner}
F. Wegner, Z. Phys. B {\bf 51}, 279 (1983).
\bibitem{brezin}
see also
E. Br\'ezin, D. J. Gross, and C. Itzykson, Nucl. Phys. B {\bf 235}, 24 (1984);
A. Klein and J. F. Perez, Nucl. Phys. B {\bf 251}, 199 (1985).
\bibitem{zphys}
see, for example, K. Broderix, H. Heldt, and H. Leschke, Z. Phys. B
  {\bf 68}, 19 (1987).
\bibitem{apel}
W. Apel, J. Phys. C {\bf 20}, L577 (1987).
\bibitem{perron}
O. Perron, {\em Die Lehre von den Kettenbr\"uchen, Vols. I, II}
  (Teubner, Stuttgart, 1977).
\bibitem{wall}
H. S. Wall, {\em Continued Fractions} (Van Nostrand, New York, 1948)
\bibitem{magnus}
W. Magnus, F. Oberhettinger, and R. P. Soni, {\em Formulas and Theorems
  for the Special Functions of Mathematical Physics},
  3rd enlarged ed.
  (Springer, Berlin, 1966).
\bibitem{growth-rate}
D. S. Lubinsky, H. N. Mhaskar, and E. B. Saff,
  Constr. Approx. {\bf 4}, 65 (1988), and references therein.
\bibitem{bender}
see, for example,
C. M. Bender and S. A. Orszag,
  {\em Advanced Mathematical Methods for Scientists and Engineers},
  (McGraw-Hill, Auckland, 1978)
\bibitem{axiom}
R. D. Jenks and R. S. Sutor, {AXIOM: The Scientific Computation System}
  (Springer, Berlin, 1992).
\end{references}
\end{document}